\documentclass[byrevtex,prd,showpacs,preprintnumbers,twocolumn]{revtex4}
\newcommand{\be}{\begin{equation}}
\newcommand{\ee}{\end{equation}}
\newcommand{\bea}{\begin{eqnarray}}
\newcommand{\eea}{\end{eqnarray}}
\newcommand{\nn}{\nonumber}

\newcommand{\ra}{\rightarrow}
\newcommand{\sqla}{\sqrt{\lambda}}
\newcommand{\tr}{{\rm tr}}

\newcommand{\bK}{{\bf K}}
\newcommand{\bE}{{\bf E}}

\newcommand{\cN}{{\mathcal N}}
\newcommand{\da}{\longleftrightarrow}
\begin{document}
\preprint{hep-th/0312113}
\preprint{AEI-2003-104}

\title{Multi-spin strings on $AdS_5\times T^{1,1}$ and
operators of ${\mathcal N}=1$ superconformal theory}

\author{Nakwoo Kim}
\email{kim@aei.mpg.de}
\affiliation{
Max-Planck-Institut f\"ur Gravitationsphysik\\
Albert-Einstein-Institut\\
Am M\"uhlenberg 1, D-14476 Golm, Germany\\
}
\begin{abstract}
We study rotating strings with multiple spins in the background of
$AdS_5\times T^{1,1}$, which is dual to a $\cN=1$
superconformal field theory with global symmetry 
$SU(2)\times SU(2)\times U(1)$ via the AdS/CFT correspondence.
We analyse the limiting behaviour of macroscopic strings and
discuss the identification of the dual
operators and how their anomalous dimensions should behave as
the global charges vary.
A class of string solutions we find are dual to operators in $SU(2)$
subsector, and our result implies that the one-loop planar dilatation operator
restricted to the $SU(2)$ subsector should be equivalent to the 
hamiltonian of the integrable Heisenberg spin chain.
\end{abstract}
\maketitle
\section{Introduction}
According to the AdS/CFT correspondence \cite{review}, the quantum
string spectrum in anti-de Sitter (AdS) space is identical to that
of a certain conformal field theory (CFT) formulated on its
boundary. The direct check of this conjecture is plagued by the
difficulty of superstring quantization in curved backgrounds so it
has been a challenge to perform a quantitative test beyond the
supersymmetric subsector and their protected data.

A way to get around this problem was suggested recently and proved
to be very successful. One makes use of classical string solitons,
not necessarily supersymmetric, and performs semiclassical
quantization of the string theory to compare with the computations
in the dual gauge theory. The inverse of a large global charge
plays the role of a new expansion parameter, enabling perturbative
computations on both sides of the duality. The celebrated BMN
limit \cite{bmn} considers small deformations of half-BPS
operators with large conformal dimensions, which are dual to
pointlike strings orbiting in the $S^5$. It amounts to taking the
Penrose limit of the given backgrounds and by exploiting the fact
that the string theory in the plane-wave becomes free in the
light-cone gauge \cite{metsaev}, one obtains all-loop results for
a large class of operators in the planar limit of the dual
conformal field theory.

This program can be extended to other classical
string solutions. The implication of the macroscopic spinning
string solutions on the dual conformal field theory was first
discussed in \cite{gkp}, and further generalized and refined
in many subsequent publications 
\cite{extension}, and 
see \cite{Tseytlin:2003ii} for a review and more complete
list of references.
More precise
comparison of the spectra has been made possible thanks to the
crucial observation that the one-loop dilatation operator
for pure scalar operators is isomorphic to the integrable
$SO(6)$ spin chains \cite{minzar}.
The anomalous dimensions for very long operators can be computed by solving
the Bethe ansatz equations in the thermodynamic limit,
which are then compared to the energy of string solitons and impressive
quantitative agreements have been observed 
\cite{agree}.
There have been
efforts to directly relate the quantum spin chain models with the
string nonlinear sigma model,
by comparing the higher conserved charges \cite{glebmatt},
or by considering the continuum limits of the
spin chains
\cite{Gorsky:2003nq,Kruczenski:2003gt},
to derive the nonlinear sigma model.

A natural question then is how much of the above developments,
especially the agreement at the quantitative level, can be
extended to other examples of the AdS/CFT correspondence, which
typically have less supersymmetries. In this paper we choose to
study the IIB string theory of $AdS_5\times T^{1,1}$, which is
dual to a $\cN=1$ superconformal field theory with $U(N)\times
U(N)$ gauge group and bifundamental matter multiplets, as first
described in \cite{kleb}. Since the string dynamics in $AdS_5$
should be identical to the maximally supersymmetric case, we
restrict ourselves on the strings moving in the squashed sphere,
$T^{1,1}$.

We stress that unlike the maximally supersymmetric case of
$AdS_5\times S^5$, the isometry of $T^{1,1}$, $SU(2)\times
SU(2)\times U(1)$, is {\it not} a consequence of the
supersymmetry. It renders the nature of our analysis highly
dynamical. As $T^{1,1}$ is mapped to the moduli space of the gauge
theory, the spinning strings in $T^{1,1}$ are dual to pure scalar
operators. We will find the on-shell relations between the
conformal dimensions and the global charges expressed implicitly
in terms of elliptic integrals. We should remark that some
spinning string solutions of $AdS_5\times T^{1,1}$ have been
studied already in \cite{Schvellinger:2003vz}. In this paper more
solutions are covered and we also discuss the dual operator to
each rotating string solution. We hope our results can direct the
gauge theory computation towards the dynamical confirmation of the
generalized AdS/CFT correspondence.

This article is organised as follows. In Sec. 2 we will briefly
review the duality of $AdS_5\times T^{1,1}$ and the dual ${\cal N}=1$
superconformal field theory. In Sec. 3 we present the
multi-spin string solutions in $T^{1,1}$.
In Sec. 4 we consider the limiting cases
when the conserved quantities become large, and discuss the
dual operators of $\cN=1$ conformal field theory.
In Sec. 5 we end with discussions and concluding remarks.
The conventions and basic properties of elliptic integrals which
were used in this paper can be found in the appendix.
\section{$T^{1,1}$ and the dual ${\cN}=1$ superconformal field theory}
In this paper we will study strings moving in $T^{1,1}$, which is
a homogeneous space $(SU(2)\times SU(2))/U(1)$, with $U(1)$ chosen
to be a diagonal subgroup of the maximal torus in $SU(2)\times
SU(2)$. The explicit form of the metric is best written as a
$U(1)$ bundle over $S^2\times S^2$, \bea ds^2 &=& a (d\theta_1^2 +
\sin^2 \theta_1 d\phi_1^2 +
  d\theta_2^2 + \sin^2 \theta_2 d\phi_2^2)
\nonumber\\
&+& b( d\psi + p\cos\theta_1 d\phi_1 + q\cos\theta_2 d\phi_2)^2 \;
, \label{metric} \eea where $\theta_i,\phi_i$ are the coordinates
of two $S^2$, and the $U(1)$ is denoted by $\psi \in [0,4\pi]$.
The space is an Einstein manifold if
$a=\frac{1}{6},b=\frac{1}{9}$, and if we further choose $p=q=1$
the space becomes supersymmetric: $T^{1,1}$ provides the angular
part of a singular Calabi-Yau manifold. One can easily see from
Eq.~(\ref{metric}) that the isometry is $SU(2)\times SU(2)\times
U(1)$. The three mutually commuting Killing vectors can be chosen
as $\partial_{\phi_1},\partial_{\phi_2},
\partial_\psi$.

The dual conformal field theory with $\cN=1$ supersymmetry,
as identified in \cite{kleb}, has gauge group $U(N)\times U(N)$,
and two chiral multiplets $A_i$ in $(N,\overline{N})$ and
another two, $B_i$, in $(\overline{N},N)$.
This theory obviously has $SU(2)\times SU(2)$ global symmetry
which act separately on the doublets $A_i,B_i$,
and also an anomaly-free $U(1)$ R-symmetry.
Altogether, these global symmetries are to be identified with
the isometry group of $T^{1,1}$.
The theory is also equipped with a quartic superpotential which is
invariant under the global symmetries,
\be
W =  \frac{g}{2} \epsilon^{ij} \epsilon^{kl}  \tr A_i B_k A_j B_l .
\ee
Combined with the conformal invariance we see that the conformal
dimension of $A_i,B_i$ should be $3/4$ at the conformal fixed point.

In this paper we are interested in the closed strings rotating in
$T^{1,1}$ with radius $\lambda^{1/4}$. As well known, the AdS/CFT
corresponde relates $\lambda$ to the 't-Hooft coupling constant of
the dual gauge field theory. We choose to work with the Polyakov
action in the conformal gauge.  We choose the ansatz that the
string is spinning along the three commuting Killing directions,
and at rest along all other directions. \bea t  =  \kappa \tau  , \;\;
\phi_1 = \omega_1 \tau , \;\; \phi_2 = \omega_2 \tau , \;\; \psi =
\nu \tau  , \nonumber \eea and the remaining two angles,
$\theta_1,\theta_2$ depend only on $\sigma$. Then it is easy to
show that the gauge fixing constraint $\dot{X} X' = 0$ is
trivially satisfied, while $\dot{X}^2 + X'^2 = 0$ becomes \bea
\kappa^2 & = & a (\theta'^2_1 + \theta'^2_2 + \omega_1^2 \sin^2
\theta_1 + \omega_2^2 \sin^2 \theta_2 )
\nn\\
&+ & b ( \nu + \omega_1 \cos \theta_1 + \omega_2 \cos \theta_2 )^2
\; ,
\label{kapp}
\eea
which is just an integrated form of the
equations of motion for $\theta_i$. Clearly we can treat the
reduced string equations of motion as a classical mechanics system
with {\it time} $\sigma$, and $\kappa$ can be identified as the
energy. We also have the periodicity condition $\theta_i (\sigma +
2\pi) = \theta_i (\sigma)$, up to the periodicity of the
coordinates $\theta_i$.

The conserved quantities we are interested in are
the energy $E  =  \sqrt{\lambda} \kappa$,
in addition to the following three angular momenta,
\bea
J_A & \equiv  & P_{\phi_1}
\nonumber\\
&=&
\sqrt{\lambda}
\int \frac{d\sigma}{2\pi}
\Big[
\omega_1 (a \sin^2 \theta_1 + b \cos^2 \theta_1)
\nonumber\\
&& \mbox{} +  b ( \nu + \omega_2 \cos \theta_2 ) \cos\theta_1
\Big] \, ,
\\
J_B &\equiv & P_{\phi_2}
\nonumber\\
&=&
\sqrt{\lambda}
\int \frac{d\sigma}{2\pi}
\Big[
\omega_2 (a \sin^2 \theta_2 + b \cos^2 \theta_2)
\nonumber\\
&& \mbox{} + b ( \nu + \omega_1 \cos \theta_1 ) \cos\theta_2 \Big]
\, ,
\\
J_R &\equiv  & P_\psi
\nonumber\\
&=& \sqrt{\lambda} \int \frac{d\sigma}{2\pi} \; b ( \nu + \omega_1
\cos \theta_1 +  \omega_2 \cos \theta_2 ) \, , \eea
which are
identified with the conformal dimension and other global charges
of the dual operators.

From what we have discussed
so far, one can easily infer the following {\it dictionary}
which is crucial in the identification of dual operators to
string solutions.
\bea
J_A & \da & \frac{1}{2}
\Big[
\#(A_1)- \#(A_2) + \#(\overline{A}_2) -\#(\overline{A}_1)
\Big]
\\
J_B & \da & \frac{1}{2}
\Big[
\#(B_1)- \#(B_2) + \#(\overline{B}_2) -\#(\overline{B}_1)
\Big]
\\
J_R & \da & \frac{1}{4}
\Big[
\#(A_i)+\#(B_i) - \#(\overline{A}_i)-\#(\overline{B}_i)
\Big]
\eea
where for instance $\#(A_1)$ counts how many times $A_1$ appears
in the dual composite operator.

In order to find general class of solutions with nontrivial
$\theta_1,\theta_2$ we need to know a constant of motion other than
the energy. Whether it is possible or not is related to the question
of integrability.
Although this is certainly a very important issue, in this paper we
consider solutions where only one coordinate, say $\theta_1$, is
activated. The one-dimensional system is readily integrated,
and it will be seen that we still have a rich class
of nontrivial solutions whose dual operators are in general non-holomorphic
combinations of the scalar fields $A_i,B_i$.

\section{Spinning Strings in $T^{1,1}$}
\subsection{Single-Spin Solutions}
Let us first consider the strings shrink to a point and orbiting
with light velocity. It can be easily seen as the solution of the
mechanical model when the particle is at rest, i.e. $\theta_i$ are
fixed at $0$ or $\pi$. If we evaluate the conserved quantities
for instance at $\theta_i=0$, we
get the simple relation 
\be b E^2 = J_A^2 = J_B^2 = J_R^2 = \lambda
b^2 (\nu + \omega_1 + \omega_2)^2 
\label{bps}
\ee 
i.e. all conserved
quantities are equal up to sign, which is determined by the values
of $\theta_i$. The linear relation between the charges implies
that this string solution is in fact supersymmetric and dual to a
chiral primary operator, which will be discussed in more detail in
the next section. We remark that the semiclassical treatment of
the string theory around the pointlike strings amounts to taking
the Penrose limit of $AdS_5\times T^{1,1}$, which was studied in
\cite{t11pp}.

Another class of simple solutions include
rotating circular strings embedded in one of the
two-spheres when $\omega_1=\omega_2=0$ and $\nu \neq 0$:
\be
\theta_1 = n \sigma , \quad\quad \theta_2 = 0 \;\;{\rm or}\;\; \pi,
\ee
in which case we have $J_A=0, J_B^2=J_R^2=\lambda b^2\nu^2$, and the
energy is related to the angular momentum as follows:
\be
E^2 = 9 J_R^2 + \lambda\frac{n^2}{6}
\, , 
\label{1circle}
\ee
These solutions can be said to be single-spin solutions,
as there is essentially only one nonvanishing component of spin.

Now we move to more general class of solutions with two spins,
and in the following we will use $\theta\equiv\theta_1,\omega\equiv\omega_1$
and set $\theta_2=0,\omega_2=0$ to simplify the notations.

\subsection{Multi-Spin Solutions}
The spinning string ansatz has been reduced to a 
one dimensional system with the following potential,
\be
V(\theta) = a \omega^2 \sin^2 \theta
+ b (\nu + \omega \cos\theta )^2 \, ,
\label{potential}
\ee
It is straightforward to integrate the equation for
generic values of $\omega,\nu$, but before we present
the full result below let us first consider the simpler
case of $\nu=0$.

For $\nu=0$, the potential has the maximal value at $\theta=\pi/2$
and without losing generality we can consider strings centered
around the north pole. When $y \equiv \kappa/\omega < \sqrt{a}$ we have
a folded string, and the relevant integral is easily transformed
into a complete elliptic integral, and from the periodicity
condition $\theta(\sigma+2\pi)=\theta(\sigma)$ we get 
\be 
\omega =
\frac{2}{\pi} \sqrt{\frac{a}{a-b}} {\bf K} 
\left( \frac{y^2-b}{a-b} 
\right)
\, . 
\ee 
It is also straightforward to express the nonvanishing
components of angular momenta as functions of $y$, 
\bea
\frac{E}{\sqla} &=& \frac{2y}{\pi} \sqrt{\frac{a}{a-b}} \, {\bf K}
\left( \frac{y^2-b}{a-b} \right) \, , 
\label{n0ef}
\\
\frac{J_A}{\sqla} &=& \frac{2}{\pi}
\sqrt{\frac{a}{a-b}}
\Big[
a \, {\bK}
\left(
\frac{y^2-b}{a-b}
\right)
\nonumber\\
&& \mbox{}-(a-b) \, {\bE} \left( \frac{y^2-b}{a-b} \right) \Big] \, , 
\\
\frac{J_B}{\sqla} &=& \sqrt{\frac{a}{a-b}} \, b \, ,
\eea 
and $J_R=J_B$, which in fact holds for any solution considered
in this section. 
The above expressions are valid for $y^2 < a$, and for larger values of $y$
the range of $\theta$ is not restricted and the string starts to
wrap the circle parametrized by $\theta$ completely. We call this
class of solutions as the circular string. The result is again
summarised in terms of the complete elliptic integrals. 
\bea 
\frac{E}{\sqla}
&=& \frac{2y}{\pi} \sqrt{\frac{a}{y^2-b}} \, {\bf K} \left(
\frac{a-b}{y^2-b} \right)
\, , 
\label{n0ec}
\\
\frac{J_A}{\sqla} &=&
\frac{2}{\pi}
\sqrt{\frac{a}{y^2-b}}
\Big[
y^2 \,
\bK
\left(
\frac{a-b}{y^2-b}
\right) 
\nonumber\\
&& \mbox{} - (y^2-b) \, \bE
\left(
\frac{a-b}{y^2-b}
\right)
\Big]
\, , 
\\
\frac{J_B}{\sqla} &=& 0 \, .
\eea

For generic values of $\nu$,
the allowed range of $\theta$ is determined from
the value of $\kappa$ compared to the maximum
value of $V(\theta)$. For $\kappa^2 < V_{\rm max}$
the string takes the shape of a folded arc, around
the north-pole or south-pole, depending on the
values of $\omega,\nu$.
From the equation of motion and the periodicity
condition $\theta(\sigma+2\pi)=\theta(\sigma)$ we get
\be
2\pi \omega  =
\sqrt{\frac{a-b}{a}}
\int \frac{d\theta}{\sqrt{(\cos\theta - \alpha)(\cos\theta - \beta)}}
\, ,
\ee
where $\alpha,\beta$ are the two roots of the quatratic equation
from the equation of motion, which we quote here for easy reference,
\be
\frac{1}{a-b}
\Big\{ b x \pm \sqrt{abx^2 - (a-b)y^2 +a(a-b)} \Big\}
\, , 
\label{roots}
\ee
and we choose $\alpha (\beta)$ to be the larger (smaller) one.
We defined $x\equiv \nu/\omega$.

Folded string requires that this quadratic equation should have
real roots, so we have the condition
\be
y^2 \leq a + \frac{ab}{a-b}
x^2
\, .
\ee

The integral becomes simpler when $\nu=0$ and reduces to the
solutions we have already discussed. Fortunately the integral can be
facilitated in terms of elliptic functions also for $\nu \neq 0$,
using the formulas presented in the appendix. It is thus possible
to express the energy in terms of $x,y$, and the integrals leading
to angular momenta can also be done. For definiteness we here consider
folded strings centered around the north pole. Strings spinning around
the south pole can be covered by considering both signs for $x\equiv
\nu/\omega$, as evident from the invariance of Eq.~(\ref{potential})
under $\theta\ra\pi-\theta,\nu\ra -\nu,\omega\ra\omega$.
\bea
\frac{E}{\sqla} &=& \frac{4y}{\pi} \sqrt{\frac{a}{a-b}}
\frac{1}{\sqrt{(1+\alpha)(1-\beta)}} \bK ( t )
\, , 
\\
\frac{J_A}{\sqla} &=&
\frac{2}{\pi}\sqrt{\frac{a}{a-b}}\frac{1}{\sqrt{(1+\alpha)(1-\beta)}}
\Big\{
\nonumber\\
&& + 2 \left[ a - bx +  (a-b)\beta \right] {\bf K}(t)
\nonumber \\
&&
+ (b-a)(1+\alpha)(1-\beta) {\bf E} (t)
\nonumber\\
&&+2 \left[ (\alpha+\beta)(b-a)+ 2 b x \right]
{\bf \Pi}(k,t)
\Big\}
\, , 
\\
\frac{J_B}{\sqla} &=&
\frac{4}{\pi}\sqrt{\frac{a}{a-b}}\frac{b}{\sqrt{(1+\alpha)(1-\beta)}}
\Big\{
\nn\\
&&
+(x-1) {\bf K}(t)
+ 2 {\bf \Pi} (k,t)
\Big\}
\, , 
\eea
where we have defined
\bea
t&=&\frac{(1-\alpha)(1+\beta)}{(1+\alpha)(1-\beta)}
\, , 
\\
k& =& -\frac{1-\alpha}{1+\alpha}
\, .
\eea

When $y$ gets larger the quadratic equation does not
have real roots, and the string is not of folded form any more
and can completely wrap the two-sphere. In the evaluation
of conserved charges the difference is that we now integrate
over $0<\theta<2\pi$, not in the range dictated by the roots
of the quadratic equation. For general circular strings the
conserved quantities are expressed as real parts of the
complete elliptic integrals evaluated
at complex numbers since the roots $\alpha,\beta$ become
complex numbers.


\section{Large Energy Limits and SCFT Operators}
We have presented several different classes of solutions
in the last section. Now let us consider the limit of
large conserved charges so that the classical relations
can be a good approximation to the quantum strings we
are eventually interested in. 

We also want to identify
the dual operators of $\cN=1$ superconformal field theory,
so that our results can be checked against gauge theory
computation in the future. Strings moving in $T^{1,1}$
are dual to pure scalar operators, which should not
contain fermions, covariant derivatives or gauge field strengths. 
The scalar fields in the dual theory are in the bi-fundamental
representations, so in order to construct gauge singlets
the fields in $(N,\overline{N})$ and in $(\overline{N},N)$
should appear alternatively. Most generally they
take the form as 
\be
\tr 
( AB \ldots A\bar{A} \ldots \bar{B}B \ldots 
\bar{B}\bar{A} \ldots
)
\, .
\ee
We note that there exists an inequality between the
bare dimension and the $R$-charge, when 
written in terms of the string variables,
\be
E \geq 3| J_R |
\, .
\label{bound}
\ee
It is just the unitarity bound which results from the
$\cN=1$ superconformal algebra.

The simplest class of solutions are the pointlike
strings, satisfying Eq.~(\ref{bps}). 
In that case we find that the energy is exactly
proportional to the angular momenta, implying protected
state. Indeed, from the relevant quantum numbers we have,
if we assume $J_R >0$, the pointlike string at $\theta_1=\theta_2=0$ is
identified as
\be
\tr (A_1 B_1 )^{2J_R}
\, , 
\ee
and for instance when at $\theta_1=\theta_2=\pi$, as
\be
\tr (A_2 B_2 )^{2J_R}
\, .
\ee
For different signs of angular momenta we can also easily
identify them as different chiral primaries.
Their bare dimensions are indeed $\frac{3}{4}\cdot 2J_R=3J_R$, 
consistent with the relation Eq.~(\ref{bps}). 
Of course these long chiral primaries constitute the
ground state of the plane-wave background as
the Penrose limit of $AdS_5\times T^{1,1}$, considered
in \cite{t11pp}.

The first non-supersymmetric example comes from the single-spin
circular strings. When we expand Eq.~(\ref{1circle}) 
for large values of the spin,
\be
E =
3 J_R + \lambda \frac{n^2}{36 J_R} - \lambda^2 
\frac{n^4}{7776 J_R^3} + \cdots
\label{1cexp}
\ee
And from the fact that $J_A=0, J_B=J_3$
we can identify the dual operator as
\be
\tr \left( (A_1 B_1)^{J_R}  (A_2 B_1)^{J_R}
+{\rm permutations} \right)
\, , 
\label{1cop}
\ee
whose bare dimension $4J_R \cdot \frac{3}{4}=3J_R$
agrees with the leading part of
Eq.~(\ref{1cexp}), and since the correction 
terms are given as a regular series expansion in
$\lambda$, we expect the subleading terms can be
checked against perturbative gauge theory 
computations in the large $N$ limit.

We remark here that the task of identifying the dual
operators for single-spin strings has been greatly simplified 
by the fact that the unitarity bound Eq.~(\ref{bound}) 
is saturated asymptotically. $E=3J_R$ first implies
that among the four combinations $AB, A\bar{A}, \bar{B}B,
\bar{B}\bar{A}$ in $(N,N)$ representations only $AB$'s
should be included. In other words, the dual operators
are holomorphic. Then $J_B=J_R$ tells us that between
$B_1$ and $B_2$, only $B_1$'s
should be used to construct the dual operator.
Finally the filling fraction of $A_i$'s are governed
by $J_A$. We also note that the conserved charges 
related to the remaining $SU(2)\times SU(2)$ generators
vanish, which means in Eq.(\ref{1cop}) we are instructed
to choose singlets of $SU(2)$ concerning $A_i$'s.
Different values of $n$, of course should denote
different eigenvectors of the same quantum numbers.

Now let us consider the folded and circular strings 
with $\nu=0$.
From Eq.~(\ref{n0ef}) and Eq.~(\ref{n0ec}) we see that in the limit
$y^2\ra a$, both $E$ and $J_B$ become large as
the complete elliptic integral of the first kind, $\bK$,
develops a logarithmic divergence. 
Obviously our strategy is to expand $E,J_A$
in terms $y^2-a$, invert the expression for $J_A$, and
substituting back into $E$ to get a relation between
$E$ and $J_A$.

For that purpose it turns out convenient to employ
so-called $q$-series expressions of elliptic integrals.
One first defines $K'(m) \equiv K(1-m)$ then for
small $m$ we have a regular series expansion for $q$,
\bea
q &\equiv& \exp \left[ -\pi K'(m)/K(m) \right]
\nonumber \\
&=&
\frac{m}{16}
+8\left(\frac{m}{16}\right)^2
+84\left(\frac{m}{16}\right)^3
+ \cdots
\eea
and for $q$-series expansion of other elliptic
integrals see the appendix.

For folded strings, we define
\be
m = \frac{a-y^2}{a-b}
\ee
and for small $m$ we have the following $q$-expansions,
using the actual values of $a,b$,
\bea
\frac{E}{\sqla} &=&
-\frac{1}{\sqrt{2}\pi}
\ln q
\Big(
1 + \frac{4}{3}q
+ \frac{100}{9}q^2
+ \cdots
\Big)
\\
\frac{J_A}{\sqla} &=&
-\frac{2\sqrt{3}}{\pi}
\Big[
\ln q
\Big(
\frac{1}{12}+\frac{1}{9}q+\frac{7}{9}q^2
+\cdots
\Big)
\nonumber\\
&&
+\Big(
\frac{1}{18}-\frac{2}{9}q
+\frac{2}{3}q^2
+\cdots
\Big)
\Big]
\eea
For circular strings, we define instead
\be
m = \frac{y^2-a}{y^2-b}
\ee
then the conserved quantities are
\bea
\frac{E}{\sqla} &=&
-\frac{1}{\sqrt{2}\pi}
\ln q
\Big(
1 - \frac{4}{3}q
+ \frac{100}{9}q^2
+ \cdots
\Big)
\\
\frac{J_A}{\sqla} &=&
-\frac{2\sqrt{3}}{\pi}
\Big[
\ln q
\Big(
\frac{1}{12}-\frac{1}{9}q+\frac{7}{9}q^2
+\cdots
\Big)
\nonumber\\
&&
+\Big(
\frac{1}{18}+\frac{2}{9}q
+\frac{2}{3}q^2
+\cdots
\Big)
\Big]
\eea
In the limit of very massive strings the two different class of
solutions exhibit very similar behaviour, both giving
\be
E =
\sqrt{6} J_A + 
\frac{\sqrt{2}}{3\pi}\sqla+
\frac{4\sqrt{6}}{27}
J_A
e^{-4\sqrt{3}\pi J_A/\sqla}
+ \cdots
\, , 
\label{nonhol}
\ee
and the difference shows up in the sub-leading terms.

Now let us consider the dual operator, on the assumption 
that in the large $E,J_A$ limit the leading behaviour
$E\sim \sqrt{6}J_A$ is given by the bare conformal dimension.
But in this case we find it is not straightforward to identify
the dual operators. The crucial difference is that the on-shell 
relation Eq.~(\ref{nonhol}) 
does not approach the unitarity bound Eq.~(\ref{bound}). 
As there are 4 scalar fields $A_i,\overline{B}_i$ in 
$(N,\overline{N})$ representation and their complex conjugates in
$(\overline{N},N)$, we have 16 different combinations in 
$(N,N)$, while we are given only 4 conserved quantities from 
string solutions.
It is clear that the identification cannot be made without
ambiguity, even up to mixing. For circular strings though,
since $J_B=J_R$ vanish identically one might conjecture that
the dual operator does not contain any $B$ or $\overline{B}$,
giving the form $\tr (A\overline{A})^{2E/3}$. Then
$E/J_A \ra \sqrt{6}$ can be used to decide
\be
\frac{
\#(A_2)+\#(\overline{A}_1)
}
{
\#(A_1)+\#(\overline{A}_2)
}
=
\frac{11-4\sqrt{6}}{5}
\, , 
\ee
But one should bear in mind that it is also possible to
consider more general forms of operators which are 
singlets of the $SU(2)$ concerning $B_i$'s.

Now let us consider the general folded string solutions.
We are again interested in the limits where the conserved quantities,
especially the energy, become very large. One can see that
here with general solutions there is another possibility for this,
than exploiting the logarithmic divergence of elliptic
integrals. The energy becomes large when $y,-x\ra\infty$, while 
keeping $t,k$ finite.
Since the expansions do not involve logarithms, we expect a
regular series expansion of $E$ in terms of $J_R$.
In order for folded strings to exist, at least one of the real roots 
in Eq.~(\ref{roots}) should reside in $[-1,1]$. It is possible when
$y,x$ are related asymptotically as
\be
y^2 = b ( x^2 + 2 c x + \cdots )
\, , 
\ee
for $-1<c<1$. We can expand the charges in terms of $1/x$, invert
it for $J_B$, and substitute back into the expressions for
$E,J_A$. We will get series expansions of the following form,
\bea
E &=& a_1 J_B + a_2 \frac{\lambda}{J_B} + a_3 \frac{\lambda^2}{J_B^3}
+\cdots
\, , 
\nonumber\\
J_A &=& b_1 J_B + b_2 \frac{\lambda}{J_B} + b_3 \frac{\lambda^2}{J_B^3}
+\cdots
\, .
\nonumber
\eea
The coefficients of the expansion, $a_n,b_n$ are determined by the
parameters of $y(x)$. The first few coefficients turn out to
depend only on the leading order correction term, $c$,
\bea
a_1 &=& 3
\, , 
\\
a_2 &=& \frac{4}{9\pi^2} {\bf K}(-z) \left( {\bf E}(-z)-{\bf K}(-z) \right)
\, , 
\label{ff0}
\\
b_1 &=& 1 - \frac{2}{z+1}\frac{{\bf E}(-z)}{{\bf K}(-z)}
\, , 
\label{ff}
\eea
where we define
\be
z =  \frac{1+c}{1-c}
\, .
\ee

We notice from $a_1=3$ that the unitarity bound is asymptotically saturated,
so the dual operators are holomorphic, and from $J_B=J_R$ it is
clear that $B_2$'s should not be included. The dual operator thus takes the
following form,
\be
\tr 
\left(
(A_1 B_1)^{2rJ_R} (A_2 B_1)^{2(1-r)J_R} + {\rm permutations}
\right)
\, .
\label{2fop}
\ee
And $b_1$ gives the filling fraction of $A_i$'s, i.e.
$b_1 = 2r-1$ and from Eq.(\ref{ff})
\be
r = 1-\frac{1}{1+z} 
\frac{{\bf E}(-z)}{{\bf K}(-z)}
\, .
\ee 
So up to the problem of mixing, we can determine the dual operators
without ambiguity, and the string soliton gives a prediction on
their anomalous dimensions.
\section{Discussions}
In this paper we have studied spinning strings in $T^{1,1}$
and obtained the on-shell relations between the conserved quantities
which can be mapped to the conformal dimensions and other
global charges of the dual operators.

Our main motivation was to find string solutions, the macroscopic limit
of which can be compared with the perturbative gauge theory 
computations. We have provided two such solutions, and the dual operators
are identified as Eq.(\ref{1cop}) and Eq.(\ref{2fop}).
A general rule can be deduced empirically from our analysis. 
The solutions which 
satisfy asymptotically the unitarity bound of the $\cN=1$ superconformal
symmetry, are dual to holomorphic operators, and are amenable to 
perturbative treatment in the gauge theory.
For the class of solutions which do not approach the unitarity bound, 
e.g. summarised as Eq.~(\ref{nonhol}), it does not seem possible
to confirm the string prediction using perturbative gauge theory
computation. And it is not the only problem: it is highly ambiguous
which gauge theory operators one should look at.

It is natural to expect that the two different types of 
asymptotic behaviour we have just observed has to do with
the supersymmetry of each solution.
We believe it would be illuminating to perform the 
$\kappa$-symmetry analysis with our solutions, 
as it was done for solutions on 
$AdS_5\times S^5$ in \cite{Mateos:2003de}.

Finding general spinning solutions
on $AdS_5\times S^5$ has been greatly facilitated
by the observation that the spinning string ansatz
leads to the integrable Neumann model, i.e. 
harmonic oscillators on the sphere \cite{neumann}.
The holomorphic operators we come across in this 
paper have excitations in only one of the $SU(2)$
structure, i.e. they do not contain $B_2$. 
We reckon that solutions with
both $\theta_1,\theta_2$ nontrivial will lead to 
more general holomorphic operators not necessarily
satisfying $J_B=J_R$.
Integrability should be a key in such an extension.

An important question is then whether the string theory
on $AdS_5\times T^{1,1}$ will turn out to be 
integrable or not. There are various indications
that it is indeed the case for $AdS_5\times S^5$, see
e.g. \cite{integrable}. 
Then the integrability of $\cN=4$ super
Yang-Mills should follow, via the AdS/CFT correspondence.
Readers are referred to Ref. \cite{Dolan:2003uh} on this issue.

To the best of our knowledge, the first hint of integrability
of strings on $AdS_5\times S^5$ comes from the classic
result of Pohlmeyer \cite{Pohlmeyer:1975nb} that hamiltonian
systems with quadratic constraints are classically 
integrable. It is well known that a conifold, i.e. a cone
over $T^{1,1}$, can be embedded
in ${\bf C}^4$ with quadratic constraints, but in order
to get the desired metric Eq.~(\ref{metric}) one considers
nontrivial K\"ahler potential. Without further analysis,
we are thus not particularly optimistic on the integrability
of the nonlinear sigma model on $T^{1,1}$.
But it is presumably worth mentioning that the one-dimensional
system we obtain using the spinning string ansatz, as 
summarised e.g. in Eq.~(\ref{kapp}), can indeed be rewritten
as a motion on ${\bf R}^4$ with flat metric and quadratic constraints.
It is certainly very desirable to check the integrability 
of nonlinear sigma models on $T^{1,1}$ and similar Einstein 
spaces relevant to Kaluza-Klein supergravity.

The integrable $SO(6)$ spin chain is reduced to 
the Heisenberg XXX$_{1/2}$ model when we
consider a subset of $SU(2)\subset SU(3)\subset SO(6)$.
When applied to the pure scalar operators of $\cN=4$
super Yang-Mills, it corresponds to operators written
in terms of two complex scalar fields.
Recently this subsector has been extensively studied in the
literature \cite{agree}, mainly because they only mix among themselves
to arbitrary orders of perturbation theory
and it is relatively easier to extend the computations
to higher orders.

In fact, the simplest nontrivial folded and circular strings on
$AdS_5\times S^5$ turned out to be dual to operators in the $SU(2)$ subsector
just mentioned, and it is observed that the leading order behavior 
extracted from the string solutions agree with the solutions of the 
Bethe ansatz equations, for a review see Ref. \cite{Tseytlin:2003ii}.

The solutions we have found are also dual to a $SU(2)$ subsector
of the $\cN=1$ superconformal field theory, as written in
Eq.~(\ref{2fop}). We here point out that the leading order data,
Eq.~(\ref{ff0}) and Eq.~(\ref{ff}), are equivalent to the
counterparts of the spinning strings on $AdS_5\times S^5$.
It can be seen by rewriting them using the identities
Eq.~(\ref{kk}) and Eq.~(\ref{ee}),
\bea
a_2 &=& \frac{4}{9\pi^2} {\bf K}(z') \left( {\bf E}(z')-(1-z'){\bf K}(z') 
\right)
\, , 
\\
b_1 &=& 1 - 2\,\frac{{\bf E}(z')}{{\bf K}(z')}
\, , 
\eea
with $z'=\frac{c+1}{2}$. They are identical to, for instance, Eq.~(7.25) and
Eq.~(7.26) of Ref. \cite{Tseytlin:2003ii}, up to overall coefficients
which can be absorbed into redefinition of the coupling constant.
We thus conjecture that the planar one-loop dilatation operator of $\cN=1$
superconformal field theory dual to $AdS_5\times T^{1,1}$, when
restricted to a $SU(2)$ subsector, is isomorphic
to the hamiltonian of Heisenberg XXX$_{1/2}$ model.
Since the full expressions governing the conserved quantities
are certainly more involved and different from that of 
spinning strings on $AdS_5\times S^5$, we presume the
universality we have observed here will be lifted
in general at higher orders of perturbation theory.

The original motivation of this work was to see
whether the integrability observed in perturbative 
$\cN=4$ super Yang-Mills theory persists with other 
nonabelian gauge theories. The tests done with the
BMN matrix model of M-theory plane-waves suggest
that integrability might be a rather general 
feature of Yang-Mills theory in the planar limit
\cite{plefka}, than naively expected.
In order to check the integrability 
beyond the $SU(2)$ subsector of $\cN=1$
example we have considered in this paper, it will be
nice to develop perturbative computations
around nontrivial conformal fixed points,
and find the effective vertex for the
dilatation operator. We hope to be able to compare
it to more general solutions of the nonlinear
sigma model on $T^{1,1}$.

\section*{Acknowledgments}
It is a pleasure to thank G. Arutyunov for illuminating
conversations and encouragement.
This work is supported by the DFG - the German Science Foundation.
\appendix
\section{Properties of Complete Elliptic Integrals}
In this appendix we present some properties of the elliptic
integrals which are used to derive the results of this note.
The complete elliptic integrals are defined as following,
\bea
{\bf K}(m) &=& \int^{\pi/2}_0 \frac{d\varphi}{\sqrt{1 - m \sin^2 \varphi}}
\\
{\bf E}(m) &=& \int^{\pi/2}_0
\sqrt{1 - m \sin^2 \varphi}
\;d\varphi
\\
{\bf \Pi}(k,m) &=& \int^{\pi/2}_0
\!\!
\frac{d\varphi}{(1- k \sin^2 \varphi)\sqrt{1 - m \sin^2 \varphi}}
\eea

The following identities immediately follow from the definitions,
\bea
{\bf K} (-m) &=&
\frac{1}{\sqrt{1+m}}
{\bf K}
\;
\Big(
\frac{m}{m+1}
\Big)
\, , 
\label{kk}
\\
{\bf E} (-m) &=&
\sqrt{1+m}
\; 
{\bf E}
\Big(
\frac{m}{m+1}
\Big)
\, .
\label{ee}
\eea 

Integrals of the following form can be expressed in terms
of the elliptic integrals,
\be
I(n) =
\int^1_\alpha \frac{s^n \, ds}{\sqrt{(s-\alpha)(s-\beta)(1-s^2)}}
\, ,
\ee
which are needed to derive the results of this note.
\bea
I(0)
&=&
\frac{2}{\sqrt{(1+\alpha)(1-\beta)}} {\bf K} (t)
\, , 
\\
I(1)
&=& \frac{2}{\sqrt{(1+\alpha)(1-\beta)}}
\Big[
2 {\bf\Pi} ( k,t) - {\bf K} (t)
\Big]
\, ,
\\
I(2)
&=&
\frac{1}{\sqrt{(1+\alpha)(1-\beta)}}
\Big[
2 (\alpha+\beta) {\bf\Pi} (k,t)
\, ,
\nn\\
&&
- 2 \alpha{\bf K} (t)
+(1+\alpha)(1-\beta) {\bf E} (t)
\Big]
\end{eqnarray}
where we have defined
\be
k = - \frac{1-\alpha}{1+\alpha} \;,
\quad\quad
t = \frac{(1-\alpha)(1+\beta)}{(1+\alpha)(1-\beta)} \; .
\ee

In order to study the elliptic integrals near the logarithmic
singularity, it is convenient to use the $q$-series,
defined as
\bea
q  & \equiv & \exp \left[ -\pi K(1-m)/K(m) \right]
\nn\\
&=&
\frac{m}{16} +
8\left(\frac{m}{16}\right)^2 +
84\left(\frac{m}{16}\right)^3
+ 992\left(\frac{m}{16}\right)^3
+ \cdots
\, .
\nn
\eea
Inverting, one obtains
\be
m
=
16( q - 8 q^2 + 44 q^3 - 192 q^4 + \cdots )
\, .
\ee
And the elliptic integrals are expressed in
$q$-series
\bea
\bK (m) & = & \frac{\pi}{2}
\left(
1 + 4q + 4q^2 + 4q^4 + \cdots
\right)
\, .
\\
\bE (m) & = &
\frac{\pi}{2}
\left(
1 - 4q + 20 q^2 + 96 q^3
\cdots
\right)
\, .
\eea

\end{document}